\documentclass[12pt]{article}

\usepackage[superscript,biblabel]{cite}

\usepackage{amsmath}
\usepackage{amsfonts}
\usepackage{amssymb}
\usepackage{graphicx}
\usepackage{graphics}
\usepackage[dvipsnames]{xcolor}
\usepackage[T1]{fontenc}

\title{\textbf{Mode Switching of Active Droplets in Macromolecular Solutions}}

\author
{Prateek Dwivedi,${}^{1}$ Atishay Shrivastava,${}^{1}$ Dipin Pillai ${}^{1}$ and Rahul Mangal ${}^{1\ast}$\\
\\
\normalsize{${}^{1}$Department of Chemical Engineering, Indian Institute of Technology Kanpur}\\
\normalsize{Kanpur, Uttar Pradesh 208016, India }\\
\\
\normalsize{$^\ast$E-mail:  mangalr@iitk.ac.in}
}

\date{}

\begin{document} 

\baselineskip24pt

\maketitle 

\renewcommand{\figurename}{FIG.}
\renewcommand{\figurename}{FIG.}

\section*{Abstract}
Typical bodily and environmental fluids encountered by biological swimmers consist of dissolved macromolecules such as proteins and polymers, often rendering them non-Newtonian. To mimic such scenarios, we investigate the motion of swimming droplets in an ambient medium doped with polymers as macromolecular solutes. Active droplets mimic the essential propulsive characteristics of several biological swimmers and serve as ideal model systems to widen our understanding of their locomotive strategies. Our experiments reveal extreme sensitivity of droplet motion to the presence of macromolecular solutes in the ambient medium. Through in-situ visualization of the self-generated chemical field around the droplet, we report unexpectedly high diffusivity of filled micelles in the presence of high molecular weight polymer solutes or macromolecules. This is attributed to the limitation of Stokes-Einstein relationship in accurately predicting micelle diffusivity due to significant size disparity between micelles and the macromolecular solute. With an increase in polymer concentration, particle image velocimetry reveals a mode-switching, from the conventional pusher mode to a puller mode of propulsion, characterized by a more persistent droplet motion. With a further increase in concentration, a secondary transition from smooth to a jittery mode of propulsion occurs. A robust Peclet number framework is proposed that successfully captures the observed mode-switching of active droplets. Our experiments unveil a novel route to orchestrate complex transitions in active droplet propulsion by doping the ambient medium with suitable choice of macromolecules.

\section*{Introduction}

The world of biological micro-organisms such as bacteria, protozoa, cilia, sperm cells, etc. is strongly governed by the interactions between individual active agents and their immediate surroundings. These motile agents are capable of sensing, responding to, and, in turn, even altering the hydrodynamic and chemical fields around them. The ambient media around the microbial organisms thus critically shape their individual as well as collective behavior. These media are often replete with diverse kinds of solutes, ranging in size from molecular to macromolecular solutes. For example, the ambient media may be imbued with macromolecular solutes such as proteins and enzymes that render them viscoelastic. Also, several bacterial species growing in aqueous media release an extracellular matrix such as exo-polysaccharides, known to be crucial for their self-organization into colonies \cite{ward1989kinetics}. While this macromolecular ambient matrix acts as a physical and chemical shield to the bacterial inmates within, it also affects the mode of propulsion characteristics, enabling quorum sensing and crucially affecting the emergent growth of colonies.

Biological micro-organisms propel themselves typically utilizing specially designed microscopic appendages attached to their surface. They consume energy from their surroundings and exhibit extreme sensitivity to external stimuli including exposure to light, external flow, physical obstruction, and chemical impurities. To mimic their unique propulsion characteristics, several recent efforts have been made to develop artificial swimmers. There are two popular kinds of artificial swimmers- Active particles (APs) and Active droplets (ADs). Self-propulsion in APs is achieved by generation of a physical/chemical gradient around the particle, through asymmetric interaction with the environment; a mechanism known as “phoresis”\cite{howse2007self}. APs typically consist of Janus colloidal particles that rely on inherent asymmetry in their surface composition to produce the desired gradient around them, and propel themselves via either self-diffusiophoresis (chemical gradient) \cite{howse2007self,golestanian2005propulsion}, self-thermophoresis (temperature gradient)\cite{palacci2013living}, or self-electrophoresis (electric potential gradient)\cite{paxton2004catalytic,brooks2019shape}. ADs, on the other hand, utilize either a chemical reaction or micellar solubilization to drive the motion of isotropic droplets of one fluid in another immiscible fluid. The surface tension gradient driven Marangoni flow is at the heart of self-propulsion in ADs\cite{izri2014self}. These artificial swimmers (APs and ADs) have been demonstrated to not only mimic microbial motion but also their response to external impetus through phenomena such as chemotaxis \cite{jin2017chemotaxis,popescu2018chemotaxis}, rheotaxis\cite{dwivedi2021rheotaxis,ranjan2020self,palacci2015artificial,dey2022oscillatory} and gravitaxis\cite{ten2014gravitaxis,castonguay2022gravitational}.

Several propulsion strategies have been proposed for making a droplet active such as  interfacial reaction\cite{thutupalli2011swarming,kitahata2011spontaneous,hirono2018locomotion}, phase separation\cite{thakur2006self}, and micellar solubilization \cite{izri2014self,kruger2016curling,dwivedi2021solute,izzet2020tunable}. Owing to its simplicity and non-reactive nature, micellar solubilization is the most popular strategy employed to generate a stable, sustained propulsion in droplets. If the surfactant concentration in the bulk is significantly higher than its critical micelle concentration (CMC), micellar solubilization spontaneously generates an interfacial tension gradient along the interface between the droplet and the surrounding ambient fluid. The resultant Marangoni stress causes the fluid at the interface to flow from the region of low interfacial tension to high interfacial tension, driving the droplet in the opposite direction. Using a convection-diffusion model for the migration of the chemical entities, Morozov and Michelin proposed that an increase in Peclet number ($Pe$) which determines the relative strength of convection in comparison to diffusion of external solutes forces the droplet to strongly interact with its own chemical field leading to a transition from a smooth motion to a jittery motion\cite{morozov2019nonlinear}. The transition was proposed to be a consequence of the onset of higher hydrodynamic modes resulting from an increase in $Pe$. Experimental studies, where either the surfactant concentration was increased\cite{izzet2020tunable} or molecular solutes were added\cite{dwivedi2021solute}, were able to validate this transition. Using a combination of experiments and linear stability theory, Suda \emph{et al.} proposed a size-dependent transition from straight to curvilinear motion\cite{suda2021straight}. However, due to considerable differences in the physicochemical complexity between different experimental systems, the precise underlying physics responsible for the transition still remains unclear. Considering the long-term applications of artificial droplets in executing intricate tasks in microfluidic domains laden with diverse chemical species, understanding the effect of the chemical complexity of the surrounding environment, in particular those which render viscoelasticity to the environment, is even more crucial. The role of macromolecular solutes in modifying microbial propulsion therefore warrants a detailed investigation.

In this work, we investigate the dynamics of active droplets in the presence of polymeric solutes with varying molecular weights and concentrations. We consider 4-pentyl-4’-cyano-biphenyl (5CB) oil droplets dispersed in an aqueous solution of tetradecyltrimethylammonium bromide (TTAB) surfactant. In this system,  the droplets have been reported to self-propel via micellar solubilization \cite{kruger2016curling,dwivedi2021solute}. A mode-switching from a smoothly propelling puller to pusher mode, followed by a transition to jittery mode, is shown to be characterized by a Peclet number, which explains observations made in the current work as well as those reported earlier \cite{dwivedi2021solute,hokmabad2021emergence,suda2021straight}. Through in-situ visualization of the trail of filled micelles in the droplet wake, we show that diffusion of filled micelles is critical in determining the onset of jittery motion.

\section*{Material and Methods}
\subsection*{Chemicals}
4-Cyano- 4’-pentylbiphenyl (5CB) was procured from Sigma Aldrich, tetradecyltrimethylammonium
bromide (TTAB) and glycerol are procured from Loba chemicals, Polyethylene Oxide of M{$_{w}$}’s ~ 35, 100, 600, 1000 and 8000 kDa were procured from Sigma Aldrich. All the chemicals were used as procured. For PIV analysis, red dyed aqueous fluro-spheres of size 500 nm were procured from Thermo Scientific. Oil soluble fluorescent dye ‘Nile Red’ was procured from Sigma Aldrich.  
\subsection*{Sample preparation}

5CB droplets were dispersed in aqueous solution of TTAB. Different samples were prepared by adding glycerol and PEO of different M{$_{w}$}’s to the aqueous solution of TTAB as solutes. 5CB droplets of size $\sim$80 µm were generated using a micro-injector (Femtojet 4i, Eppendorf) by injecting appropriate amount of nematic liquid crystal 5CB into aqueous solution of TTAB surfactant and solutes (glycerol and polymers). To ensure low number density of droplets to reduce droplet-droplet interaction, low concentration of 5CB, ~ 0.15 v/v\%  was used. The resulting emulsion was subsequently injected into a custom-made Hele-Shaw-type optical cell (vertical gap of 100 µm), prepared using cleaned (ultrasonicated in ethanol followed by plasma treated and nitrogen dried) glass slides. Fresh optical cells were used for every sample.
\subsection*{Rheology measurements}
Steady shear oscillatory rheology experiments were performed to measure the variation of bulk viscosity ($\eta_{b}$) vs shear rate ($\dot{\gamma}$) of different fluids. Polymer relaxation time ($\tau_{p}$) was obtained from the crossover of storage modulus ($G'(\omega)$) and loss modulus ($G''(\omega)$), measured using frequency sweep measurements at strain ($\gamma$) of 1\%. All measurements were performed at 25$^{\circ} C$ in a rheometer (TA Instruments DHR3) fitted with a Couette assembly.
\subsection*{Droplet visualization and tracking}
To visualize motion of isolated active droplets, the optical cell was placed on a temperature stage mounted on an upright polarized optical microscope Olympus BX53. The temperature stage was used to maintain target temperature of 25$^{\circ} C$ of the sample. Olympus LC30 camera with 2048 x 1532 pixels resolution was used in brightfield mode to capture the motion of the droplets. Droplet tracking is performed with Image-J software which utilizes an image correlation-based approach to obtain particle trajectories (X (µm), Y (µm)) vs. elapsed time ($\Delta t$)
\subsection*{Fluorescence experiments}
For observation of the chemical field around the droplet, oil soluble fluorescent dye (Nile Red) was dissolved in 5CB oil. Laser of wavelength $\sim$560 nm was used to excite the dye molecules. For obtaining hydrodynamic field around the droplet, surrounding aqueous solution was doped with red fluorescent tracers (500nm Polystyrene particles) and particle image velocimetry (PIV) experiments are performed. During the fluorescence experiments, ORX-10G-71S7C-C, FLIR, camera with 3208 x 2200 pixels was used with constant exposure throughout the video.

\section*{Results and Discussion}

For the purpose of benchmarking our experimental setup, we begin by reproducing well-established observations of active droplets ($\sim$80 $\mu$m) in an aqueous solution of TTAB, at first devoid of any solute, and subsequently with glycerol added as a molecular additive. In the absence of any solute, the droplets perform smooth ballistic motion with propulsion speeds $\sim$20-50 $\mu$m s$^{-1}$ at small times and random motion at long times. The droplet trajectories demonstrate traces of curling behaviour due to the nematic phase of the 5CB liquid crystal droplet \cite{kruger2016curling}. Since the density of 5CB ($\sim$1.057 g ml$^{-1}$) is higher than water ($\sim$0.99 g ml$^{-1}$), the droplets remain confined in the 2D (X-Y) plane close to the bottom wall of the optical cell and move with negligible vertical (Z) drift, as also observed in previous reports\cite{dwivedi2021solute,dwivedi2021rheotaxis}. Further, with the addition of glycerol as solute ($c_{solute}$= $c_{Glycerol}$= 94wt\%), consistent with recent experimental reports \cite{dwivedi2021solute,hokmabad2021emergence}, the droplets exhibit jittery motion. Although the density of glycerol ($\sim$1.25 g ml$^{-1}$) is greater than 5CB ($\sim$1.057 g ml$^{-1}$), the vertical height of the optical cell being comparable to the droplet size ensures that the droplets remain confined in the 2D plane.

Having benchmarked our experimental setup, we proceed to systematically investigate the effect of macromolecular solutes on droplet self-propulsion. To this effect, polyethylene oxide (PEO) of different molecular weights (M{$_{w}$}) is added to the aqueous surfactant solution as a macromolecular additive. The concentration of PEO of different M{$_{w}$} is tuned such that the bulk viscosity ($\eta_{b}$) of the ambient fluid is equal in all cases. It is to be noted that for all polymeric samples, $c^{*}\ll c_{PEO}\ll$ 1, suggesting a semi-dilute regime. Here, 
\begin{math}
        c^{*}= 
        \frac{3M_w}{4\pi R_{g}^{3}N_{A}}
\end{math} 
is the overlap concentration of PEO, $R_{g}$ is the radius of gyration of PEO in a good solvent (e.g. water), and $N_A$ is the Avogadro’s constant \cite{zikebacz2011crossover}. 

\subsection*{Motility in Viscous Macromolecular Media}

The addition of macromolecular solutes to the ambient medium may potentially affect the bulk viscosity ($\eta_{b}$), solute diffusivity or induce viscoelasticity in the medium. To isolate the effects of each of these, we begin with a discussion of the experiments wherein the addition of macromolecules does not introduce any noticeable elastic effects in the ambient medium, i.e., low Deborah number ($De=\frac{\tau_p \langle v \rangle}{R}\ll1$). 
Here, $\tau_p$ is the polymer relaxation time, $\langle v \rangle$ = 
\begin{math}
\langle \mathbf{v}_{inst.} \rangle=
\biggl \langle |\frac{\mathbf{r}_{i+1}-\mathbf{r}_{i}}{t_{i+1}-t_{i}} |\biggr \rangle
\end{math}
is the time-averaged droplet speed, and $R$ is the droplet size. Here, $\mathbf{r}_{i}=\{X_{i}(t), {Y}_{i}(t)\}$, is the instantaneous position vector of the droplet. The role of macromolecules in such systems is limited to altering the bulk viscosity of the ambient medium and surfactant diffusivity. The effect on surfactant diffusivity is of particular significance and shall be elaborated upon later. The composition and physical properties of the ambient medium in the presence of different solutes are provided in Table \ref{tab:fluids}. As seen from the table, Fluid 1 represents the ambient medium devoid of any solute, while Fluid 2 is 94 wt.\% of the molecular additive glycerol. Fluids 3 to 7 represent the ambient medium added with PEO of increasing molecular weight, ranging from 35 to 8000 kDa. 
\begin{table} [ht]
   \caption{Characteristics of different ambient fluids.}
    \centering
    \begin{tabular}{ c  c  c  c  c c c}
    \hline
    Fluid & Solute & M{$_{w}$} & $c_{solute}$ & $\eta_{b}$ &$\frac{dR}{dt}$ & Motion \\ [0.1ex]
   & & $kDa$ & wt$\%$ & $Pa\cdot s$ & $\mu$m min$^{-1}$\\
    \hline 
    1 & NA & NA & NA & 0.001 & 0.57 & Smooth\\
    2 & Glycerol & NA & 94 & 0.32 & 1.50 & Jittery\\
    3 & PEO & 35 & 25 & 0.29 & 0.54 & Jittery\\
    %  Water+TTAB+PEO & 6 & 8x10$^6$ & 1.0 & 11& No\\
    4 & PEO & 100 & 15 & 0.45 & 0.46 & Jittery\\
    5 & PEO & 600 & 3 & 0.40 & 0.30 & Smooth\\
    6 & PEO & 1000 & 2 & 0.37 & 0.33 & Smooth\\
    7 & PEO & 8000 & 0.5 & 0.37 & 0.32 & Smooth\\
   \hline
    \end{tabular}
    \label{tab:fluids}
    \end{table}

%%--------------figure 1----------------------------%%
\begin{figure}
\centering
\includegraphics[width=0.8\linewidth]{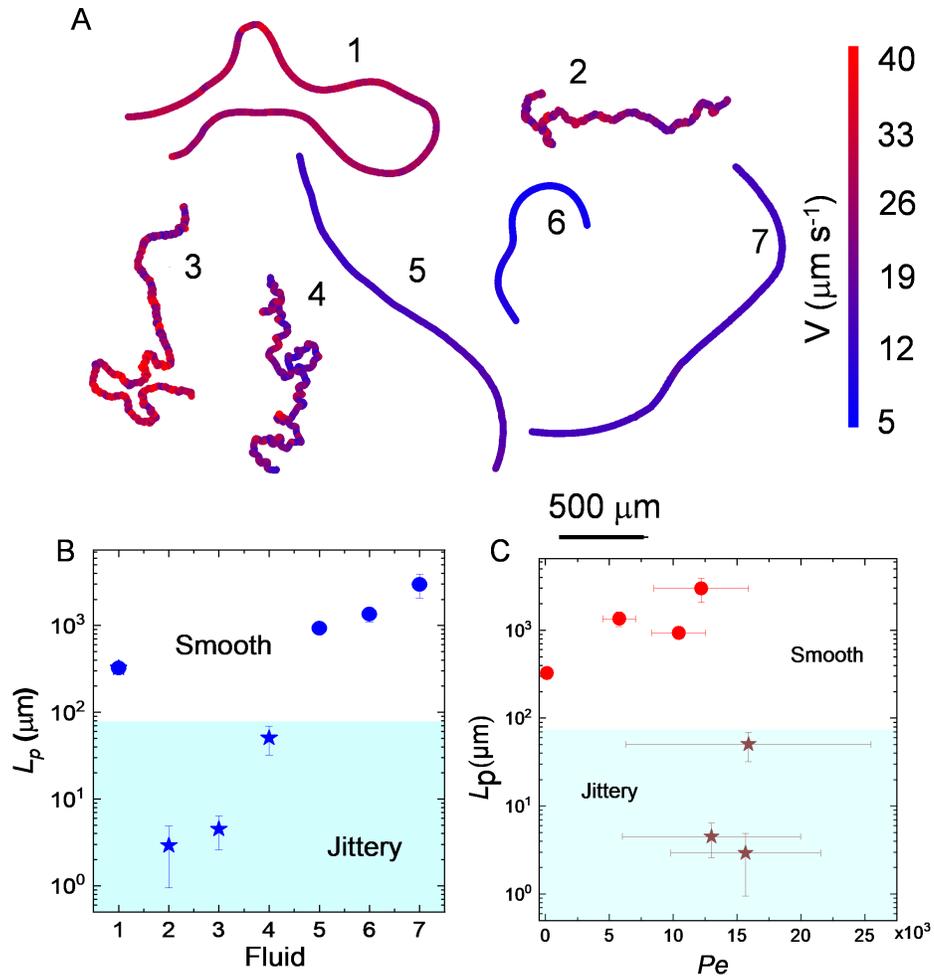}
\caption{A. Representative trajectories (for a duration of $\sim$180 s) of active 5CB droplets in different ambient media listed in table \ref{tab:fluids}. Variation in persistence length, $L_{p}$, for different ambient media. C. Variation in $L_{p}$ with Peclet number $Pe$.}
\label{control}
\end{figure}

FIG. 1A shows the typical droplet trajectories (captured for $\sim$180s) in the different ambient media listed in Table \ref{tab:fluids}. For low  M{$_{w}$}  PEO as solute (fluid 3), the motion of the active droplet is jittery and qualitatively similar to that of glycerol as solute (fluid 2). With slightly higher M{$_{w}$}  PEO (fluid 4), the jittery motion is slightly suppressed. For three further higher M{$_{w}$} PEO solutes (fluid 5, 6 and 7), the motion is smooth and similar to that in the absence of any solute (fluid 1). FIG. 1B depicts the persistence length ($L_{p}=\langle v \rangle \tau$) of droplet trajectory in different fluids. Here, $\tau$ is the rotational time scale of the droplet computed using the temporal decay of the velocity autocorrelation $C(\Delta t)=\langle \mathbf{v}_{inst.}(\Delta t).\mathbf{v}_{inst.}(0) \rangle$. Systems with large $L_{p}$ exhibit smooth droplet motion, whereas with increasing jitteriness, $L_{p}$ decreases. The bulk viscosity ($\eta_{b}$) of the different ambient media and the corresponding rate of solubilization ($\frac{dR}{dt}$) of active droplets are listed in table \ref{tab:fluids}. The bulk viscosity is determined using steady shear oscillatory rheometry (see Materials and Methods). The data in table \ref{tab:fluids} suggest that neither ($\eta_{b}$ ) nor $\frac{dR}{dt}$ in isolation correlates with the observed transition from smooth to jittery motion. This is in contradiction to the recent studies attributing the jittery behavior to either bulk viscosity \cite{hokmabad2021emergence} or the enhanced rate of solubilization \cite{dwivedi2021solute}. The theoretical model by Morozov and Michelin\cite{morozov2019nonlinear} predicts a transition from smooth to jittery motion if the Peclet number ($Pe= \frac{R V_{t}}{D}$) exceeds significantly beyond the critical threshold. They showed at high values of $Pe$, the emergence of higher modes of convection may result in a transition from steady propulsive state to a transient chaotic motion. In the above, $R$ is the droplet size and $V_{t}$ is the characteristic velocity scale given by the terminal velocity of the droplet under the combined influence of diffusiophoretic and Marangoni effects. Some studies have used $D$ as the diffusion coefficient of the surfactant monomer\cite{morozov2019nonlinear,michelin2013spontaneous}, whereas, others have reported $D$ as the diffusion coefficient of the micelles\cite{izri2014self,morozov2020adsorption}. Taking the experimentally measured average speed of the droplet ($\langle v \rangle$) as the characteristic velocity scale, and the bulk viscosity ($\eta_{b}$) to compute micelle diffusivity using the Stokes-Einstien equation, $D_{SE}=\frac{k_{B}T}{6\pi\eta_{b}a}$, we determine $Pe=\frac{R\langle v \rangle}{D_{SE}}$  for active droplets in different ambient media. Here, $k_{B}$ is the Boltzmann constant, $T$ is the temperature, $a$ is the micelle size ($\sim$5 nm), and $\eta_{b}$ is the bulk viscosity of the ambient medium. FIG. 1C shows the variation of ${L_{p}}$ with $Pe$, which reveals no correlation between  $Pe$ and the transition from smooth to jittery motion of droplets in macromolecular solutions. We show that this discrepancy may be attributed the size of macromolecular polymeric solutes affecting the diffusivity of filled micelles.

\begin{figure}
\centering
\includegraphics[width=0.8\linewidth]{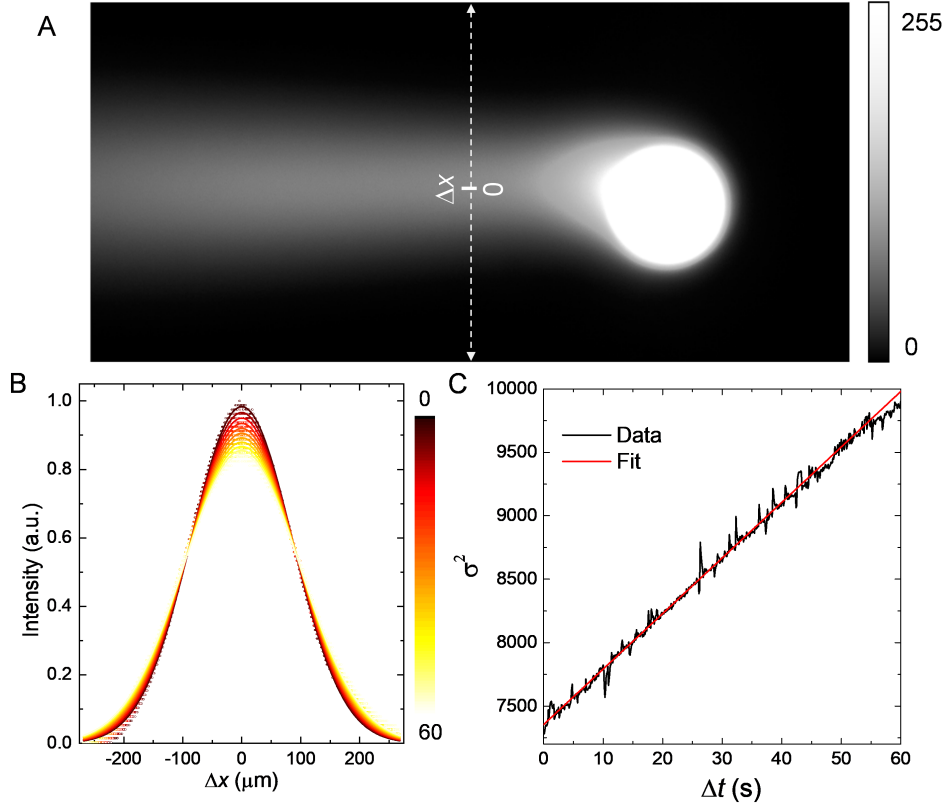}
\caption{A. Fluorescence micrograph (greyscale) of an active 5CB droplet leaving a trail of filled micelles in its wake. B. The time evolution of fluorescence intensity along the dashed line shown in panel (A) overlaid with corresponding Gaussian fits. C. Temporal change in variance ($\sigma ^2$) of the Gaussian fit shown in panel (B). The solid line represents a linear fit.}
\label{fluoro}
\end{figure}

In order to experimentally measure the micelle diffusivity, $D$, using the protocol adopted from Hokmabad \emph{et al.}, the droplet phase was doped with a hydrophobic fluorescent dye (Nile Red, Sigma Aldrich) and excited using a laser light of 560 nm wavelength\cite{hokmabad2022chemotactic}. This facilitates the visualization of the trail of filled micelles released by the droplets during their self-propulsion (FIG. 2A). The spread of this trail in the direction perpendicular to droplet motion is indicative of the diffusion of filled micelles. To avoid the oversaturated signal at the droplet, we extract the intensity $I(x,\Delta t)$ of  the fluorescence at a distance $\sim$ $4R$ from the droplet center in its wake at different times, along a line perpendicular to the droplet motion (FIG. 2B). As shown in the figure, $I(x)$ expectedly shows a Gaussian profile, which can be fitted using the Gaussian distribution:
\begin{equation}
        I(x, \Delta t)= 
        \frac{M_o}{4 \pi D \Delta t} exp(\frac{-x^2}{4D\Delta t})
        \end{equation}  
where,  \begin{equation}
        M_o= 
        \int_{-\infty}^{\infty} I(x,0) \,dx 
        \end{equation} 

The Gaussian curve-fitting suggests that the spatial variance increases linearly with time ($\Delta t$), given by $\sigma^{2}=2D\Delta t$ (FIG. 2C), whence the filled micelle diffusivity, $D$, can be determined. FIG. 3A depicts the variation of $\frac{D}{D_{SE}}$ for the filled micelle in different ambient media. The plot indicates that for polymeric solutions with large M{$_{w}$} PEO, the experimentally measured diffusivity is approximately two orders of magnitude higher than $D_{SE}$. This positive deviation in diffusivity of the filled micelles is reminiscent of enhanced diffusivity observed for nanoparticles (NPs) of the size $\sim$5-10 nm in polymeric melts/solutions, which is attributed to the significant size difference between the NPs and the  characterstic length scale of the polymer chains in the suspending medium \cite{cai2011mobility,yamamoto2015microscopic,kalathi2014nanoparticle,mangal2015phase,mangal2016size}. When the size of dispersed NPs is smaller or equivalent to the polymer length scale, the continuum hypothesis breaks down. Owing to their relative smaller size, NPs do not experience the bulk-scale drag offered by the polymer chains characterized by the bulk viscosity $\eta_{b}$ . Instead, the diffusion of NPs is mainly governed by the local viscosity ($\eta_{local}$ ), which depending on the size disparity can be significantly lower than $\eta_{b}$, even approaching the solvent viscosity, i.e., $\eta_{water}$ , in the limiting case. Since the PEO in our experiments is in the semi-dilute regime, the relevant characteristic polymeric length scales are the correlation length $\xi$ =
\begin{math}
b \phi^{-0.76},
\end{math}
and the tube diameter ($d_{t}$)=
\begin{math}
b\sqrt{N_{e}}\phi^{-0.76}
\end{math}, assuming good solvent conditions \cite{cai2011mobility,nath2018dynamics}. Here, the volume fraction,  $\phi$=$\frac{c_{PEO}}{\rho_{PEO}}$, $\rho_{PEO}$ = 1.125 g ml$^{-1}$ and $b=1.1$ nm is the length of a Kuhn segment for PEO chains. These two length scales have been listed in table \ref{tab:scales} for the different ambient media. 
\begin{figure*}
\centering
\includegraphics[width=0.8\linewidth]{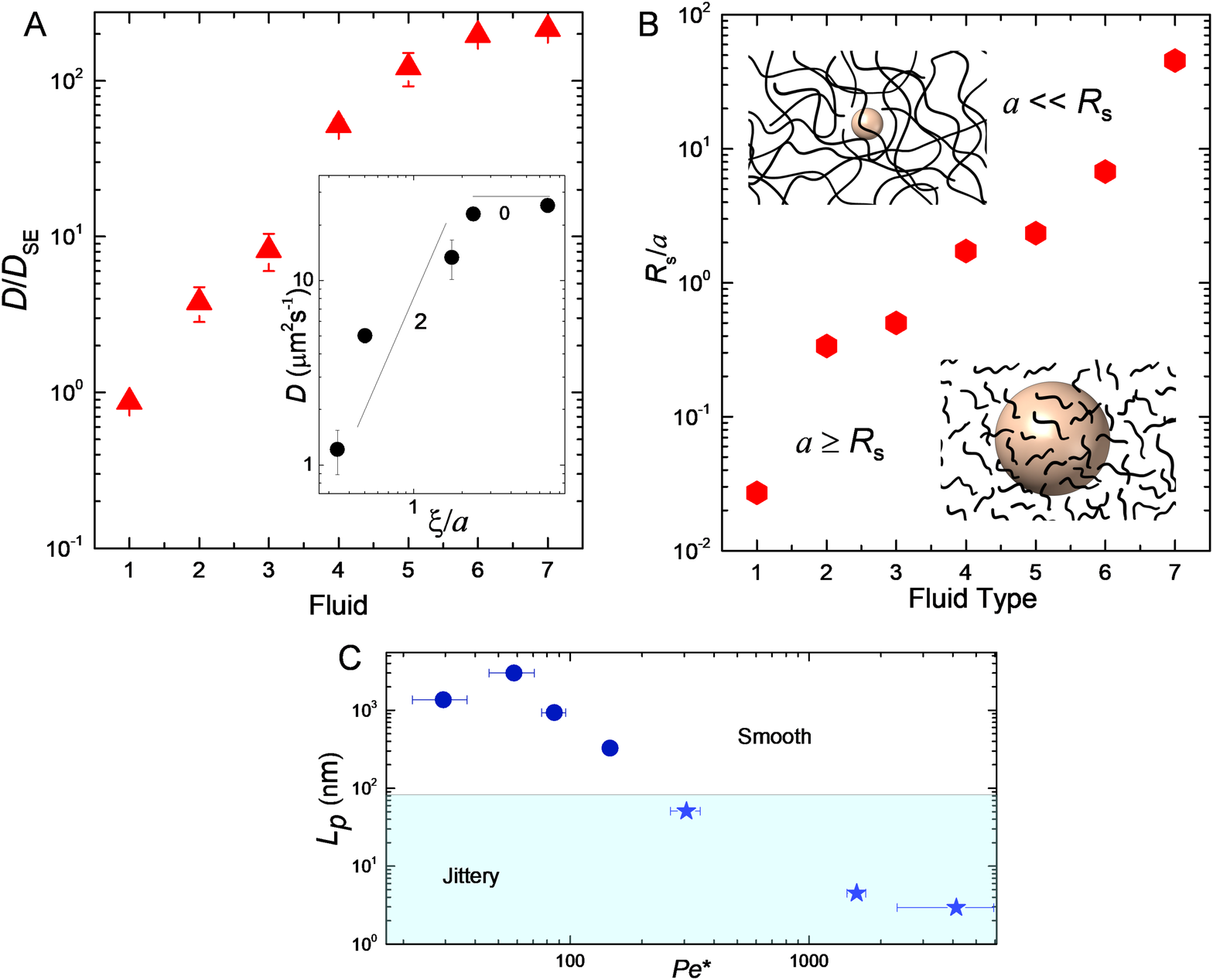}
\caption{A. Variation of $\frac{D}{D_{SE}}$ for filled micelles in different ambient media. The inset shows the variation of $D$ with $\frac{\xi}{a}$ for macromolecular solutions. B. Variation in $\frac{R_{s}}{a}$ with fluid type. The insets depict the schematic representation of micelle in the small size ($a << R_{s}$) and intermediate size range ($a > R_{s}$). C. Variation in $L_p$ with $Pe^*$.}
\label{micelle diffusivity}
\end{figure*}
Based on the scaling law proposed by Cai \emph{et al.}\cite{cai2011mobility}, we expect that micelles in the small size range ($a < \xi$) are effectively dragged by the local solvent viscosity ($\eta_{local}$ = $\eta_{water}$) suggesting that
\begin{math}
D\sim\frac{k_{B}T}{\eta_{water}a}
\end{math} . For intermediate sized micelles ($a < \xi < d_{t}$), the effective local viscosity ($\eta_{local}$) experienced by the micelles is proportional to the number of correlation blobs in a polymer chain section with size of the order of particle diameter ( \begin{math}
\eta_{local}=
\eta_{water}(\frac{\xi}{a})^{-2}
\end{math}), suggesting that
\begin{math}
D \sim
\frac{k_{B} T}{\eta_s a} (\frac{\xi}{a})^2
\end{math}. FIG. 3B shows the ratio of solute characteristic length scale to micelle size ($\frac{R_{s}}{a}$) for different fluids. Here, $R_{s}$  equals the molecular size of glycerol in the case of fluid 2, and $R_{s}=\xi$ as listed in table \ref{tab:scales} for macromolecular solutions such as fluids 3 to 7. As discussed earlier, due to the 
\begin{table} [ht]
   \caption{Characteristic length scales.}
    \centering
    \begin{tabular}{ c  c  c  c  }
    \hline
    PEO M{$_{w}$} & $c_{PEO}$ & $\xi$ &$d_{t}$  \\ [0.1ex]
    $kDa$ & wt$\%$ & (nm) & (nm)\\
    \hline 
    35 & 25 & 3.37 & 9.97 \\
    100 & 15 & 5.02 & 17.02 \\
    600 & 3 & 17.24 & 58.46 \\
    1000 & 2 & 23.48 & 79.62 \\
    8000 & 0.5 & 67.43 & 228.65\\
   \hline
    \end{tabular}
    \label{tab:scales}
    \end{table}
large size disparity between the micelles and the high M{$_{w}$} PEO solutes, we expect the continuum-based Stokes-Einstein relationship to fail, leading to a much faster micelle diffusivity. The inset of FIG. 3A depicts the variation of experimentally measured $D$ of micelles in polymeric solution with $\frac{\xi}{a}$, which is in agreement with the scaling law predicted by Cai \emph{et al.}, confirming that the motion of micelles is governed by the local viscous drag \cite{cai2011mobility}. Next, using the experimentally measured $D$ for the micelles which captures the effect of $\eta_{local}$, we compute a modified $Pe^{*}=\frac{R\langle v \rangle}{D}$. FIG. 3C displays the variation in $L_{p}$ with $Pe^{*}$, where a monotonic decrease in $L_{p}$ with increasing $Pe^{*}$ is evident, in agreement with the extant theoretical predictions \cite{morozov2020adsorption}. Our study suggests that the validity of conventional $Pe$, defined in terms of Stokes-Einstein diffusivity is limited to molecular solutes only, wherein the size of a micelle is of the same order as that of the added solute, and so the viscosity experienced by the micelle equals the bulk viscosity $\eta_b$. However, for macromolecular solutes such as polymers, due to the significant size difference between the micelles and the solutes, a modified $Pe^{*}$, where $D$ is defined using $\eta_{local}$ should be used. This accurately captures the enhanced diffusion of micelles in macromolecular solutions and therefore the jittery transition as well. In the limiting case of molecular solutes, $\eta_{local}$= $\eta_{b}$, we recover $Pe^*$ = $Pe$ , rendering $Pe^{*}$ as the universal parameter that characterizes the transition from smooth to jittery motion for all kinds of solutes, whether molecular or macromolecular.

\subsection*{Motility in Weakly Viscoelastic Macromolecular Media}
So far, we limited our discussion to Newtonian ambient media ($De$= 0), allowing us to solely investigate the effect of change in molecular weight of macrosolutes or  $\eta_{local}$, on the motion of the active droplets. To explore the effect of viscoelasticity, the motion of various $\sim$80 $\mu$m active droplets was investigated in ambient media with 8000 kDa PEO at varying concentrations ($c_{PEO}$). The relaxation time ($\tau_{p}$) and bulk viscosity ($\eta_{b}$) of these polymeric solutions were measured from the oscillatory shear rheology experiments. FIG. 4A shows an increase in $\eta_{b}$ and a consequent decrease in the droplet speed $\langle v \rangle$  with an increase in $c_{PEO}$. In these systems, since the surfactant concentration is maintained constant ( $c_{TTAB}$= 6wt $\%$), a nearly unchanged rate of solubilization  ($\sim$0.3 $\mu$m min$^{-1}$) is observed.  Since the polymer’s characteristic length scale $\xi$ ($\sim$50 nm) is greater than the micelle size ($\sim$5 nm), $D$ also remains nearly constant ($\sim$24 $\mu$m$^2$ s$^{-1}$). It is noteworthy that despite a 10$^4$ times increase of $\eta_{b}$ for $c_{PEO}$=1 wt$\%$, the reduction in droplet speed is merely $\frac{1}{10}$ relative to water. Zhu \emph{et al.} investigated  swimmers with different swimming gaits in polymeric solutions using numerical simulation to predict that polymers present in the solvent can transfer stored elastic energy due to deformation to the swimming droplet, making the swimming process more efficient in these media in comparison to Newtonian fluids\cite{zhu2012self}. With an increase in $c_{PEO}$, $\eta_{b}$ increases and thereby results in a decrease in $Pe^*$ (FIG. 4B). Despite a decrease in droplet speed $\langle v \rangle$ with increase in $c_{PEO}$, a stronger increase in $\tau_{p}$ results in an increase in $De$ (FIG. 4B). FIG. 4C demonstrates slower decay in velocity autocorrelation  $C(\Delta t)$ with increasing $c_{PEO}$ suggesting a more persistent motion in viscoelastic solutions. A comparison of few representative droplet trajectories in pure water and in 1wt$\%$ 8000kDa PEO solution (FIG. 4D) indicates that the trajectories are persistent in the latter. An increasingly persistent motion with a decrease in $Pe^*$ is also consistent with previous studies on molecular solutes \cite{morozov2019nonlinear,hokmabad2021emergence}. Using water in oil emulsion, Suda \emph{et al.} reported that upon increasing the droplet size, its motion becomes increasingly curved, indirectly suggesting that low $Pe$ is associated with more persistent motion\cite{suda2021straight}. They also reported a transition in swimming mode from puller to pusher, with increase in droplet size (and thus $Pe$), which was accompanied by straight-to-curved transition in its motion. The pusher mode was demonstrated to be more susceptible to perturbations, where the higher swimming modes become dominant leading to change in swimming direction.

\begin{figure*}
\centering
\includegraphics[width=0.8\linewidth]{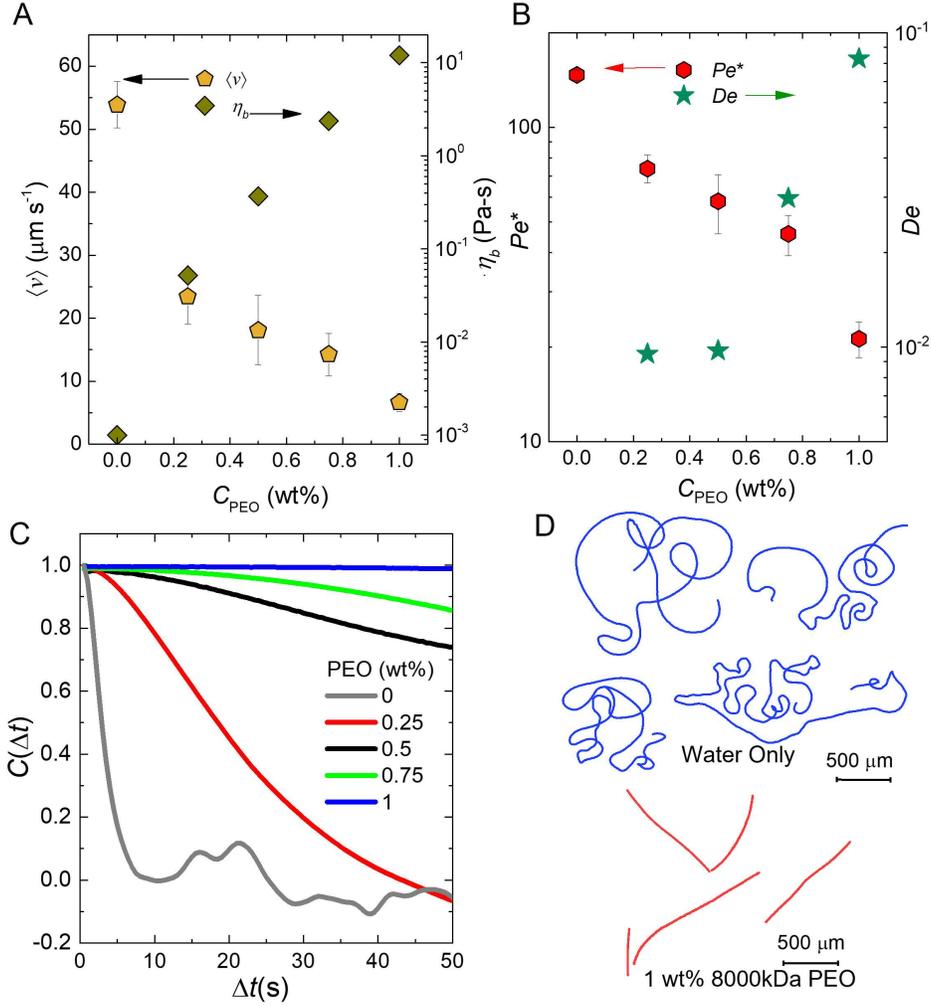}
\caption{A-B. Variation in propulsion speed $\langle v \rangle$, bulk viscosity $\eta_{b}$, $Pe^*$ and $De$ for active droplets in polymeric solutions with 8000 kDa PEO at different PEO concentrations $c_{PEO}$. C. Velocity autocorrelation $C(\Delta t)$ for different PEO concentrations ($c_{PEO}$). D. A few representative trajectories (for a duration of 200 s) comparing the persistent motion of active 5CB droplets in polymeric samples with $c_{PEO}$=0 wt$\%$ (blue) and $c_{PEO}$=1 wt$\%$ (red).}
\label{speedvsviscosity}
\end{figure*}

We also performed particle image velocimetry (PIV) to estimate the flow field around the active droplets and thereby ascertain their swimming mode (see Materials and Methods). The streamlines in the laboratory frame for fluids with $c_{PEO}$ =0 and 1.5 wt$\%$ are shown in FIG. 5(A,B). In FIG. 5A, the
fluid is being pushed away from the droplet apex, with the maximum
interfacial speed occurring in the posterior hemisphere of the droplet, thereby exhibiting a pusher mode of
swimming. In contrast, the droplet moving in 8000kDa PEO aqueous solution with $c_{PEO}$=1.5 wt$\%$ has fluid being pulled in from the front (FIG. 5B) with maximum interfacial speed occurring in the anterior hemisphere of droplet, thereby exhibiting a puller mode of swimming. The PIV data was used to compute the tangential velocity at the droplet interface $u(R,\theta)$ in the co-moving frame of reference (FIG. 5D). It can be expressed as 
\begin{math}
u(R,\theta) =
\Sigma_{n=1}^{\infty} B_n V_n (\cos{\theta})
\end{math}, where 
\begin{math}
V_n(\cos{\theta}) =
\frac{2}{n(n+1)}\sin{\theta}P_n \cos{\theta}
\end{math}, and $P_{n}$ is the Legendre polynomial of degree $n$ \cite{lighthill1952squirming,blake1971spherical}. A convenient and tractable expression is obtained by taking $B_{n}$=0 for $n>$2, resulting in 
\begin{equation}
        u(R,\theta)= 
        B_1 \sin{\theta}+\frac{B_2}{2} \sin{2\theta}
        \end{equation}
\begin{figure*}
\centering
\includegraphics[width=1\linewidth]{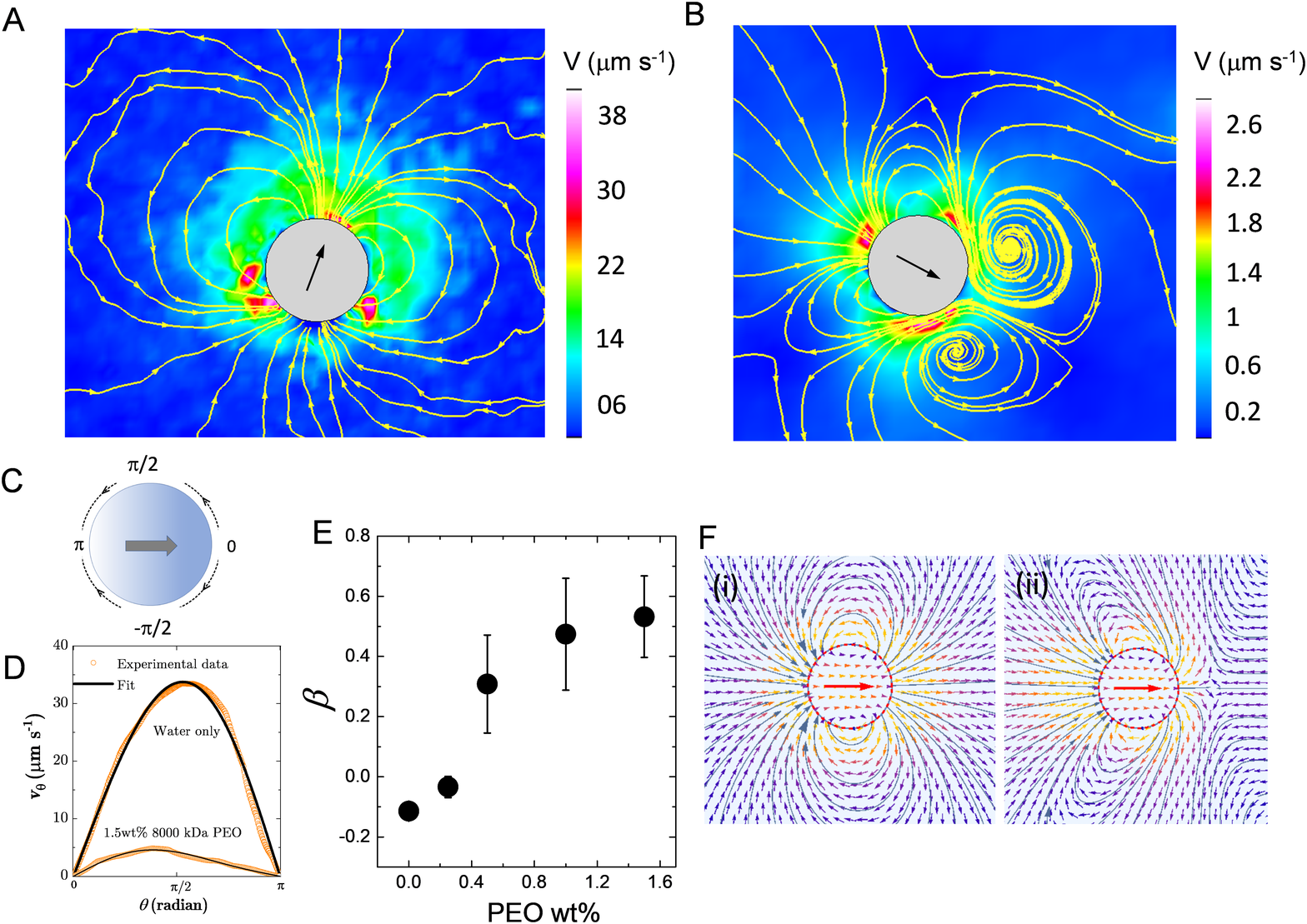}
\caption{A-B. Fluorescent micrographs depicting streamlines obtained from PIV measurements. The streamlines represent the flow-field around the droplet. A is for droplet motion in just water ($c_{PEO}$ =0 wt$\%$) where the droplet appears to be a weak pusher. B corresponds to droplet motion in ($c_{PEO}$ =1.5 wt$\%$) where it appears to be a puller.  C. Schematic of the tangential velocity field around an active droplet. D. Tangential flow velocity, obtained from PIV measurements, at the droplet interface, in the co-moving frame of reference. Black lines are theoretical fits from Eq. (3) \cite{golestanian2005propulsion}. E. $\beta$ obtained from the theoretical fitting of $u(R,\theta)$ as a function of PEO wt$\%$. F. Theoretical flow profile using the squirmer model \cite{blake1971spherical} for (i) a weak pusher active droplet with $\beta$ =-0.05, (ii) a puller active droplet with $\beta$=0.5}
\label{PIV}
\end{figure*}        
        
The ratio, $\beta$=$\frac{B_2}{B_1}$, determines the nature of swimming mode. For pusher $\beta <$0, for neutral $\beta$ =0, and for puller $\beta >$0 \cite{pedley2016spherical}. The theoretical velocity profile obtained by solving the squirmer model \cite{blake1971spherical} is depicted in FIG. 5F for two prescribed values of $\beta$ ,viz., (i) $\beta$ = -0.05 and (ii)  $\beta$ =0.5. These correspond to the values determined experimentally for the profiles shown in FIG. 5A and 5B respectively. It can be seen that the experimental and theoretical profiles are in qualitative agreement. FIG. 5E reveals an increase in $\beta$ with increasing $c_{PEO}$, marking a mode-switching from pusher to puller mode of propulsion. This novel pusher to puller transition, at first, appears to be due to an increase in the viscoelasticity characterized by an increase in $De$. However, as shown in FIG. 4B, an increase in $De$ is also accompanied by a decrease in $Pe^*$, suggesting this to be similar to an advection-diffusion related transition, reported in the literature for Newtonian media \cite{morozov2019nonlinear}. To establish this, we perform additional experiments with concentration of 4wt$\%$ and 5wt$\%$ of relatively low M{$_{w}$} of 600 kDa PEO to achieve low $Pe^*$ at very low $De$. Under these conditions, the droplets exhibit puller mode, confirming that the pusher to puller transition is governed primarily by a decrease in $Pe^*$. It is important to note here that the pusher to puller transition with decreasing $Pe^*$ has not been predicted by the extant theories/numerical simulations so far. Very recently, Li performed a direct numerical simulation and reported an unsteady droplet motion leading to zig-zag trajectories with alternating pusher–puller modes. However, this prediction was made for systems only at high $Pe^*$\cite{li2022swimming}. It was demonstrated that the interaction between the primary wake and the secondary wake of low surfactant concentration leads to this unusual swimming mode. Experimentally, pusher to puller transition for swimming oil droplets in aqueous surfactant solution has not been reported yet. For water in oil droplet system, Suda \emph{et al.} reported the existence of puller mode at low $Pe$. The underlying physics for their transition is expected to be similar to that in our system. The lack of experimental observation of a puller mode for oil droplets in an aqueous medium devoid of macromolecules can be attributed to the practical challenges in experimetally achieving low enough $Pe^*$. Our work presents a novel route of using macromolecules to increase the bulk viscosity without altering the local micelle-scale viscosity to achieve low enough $Pe^*$ with ease and explore a regime which, so far, was inaccessible for oil in water emulsions. 

We now proceed to propose a plausible hypothesis for the observed mode-switching from pusher to puller with a decrease in $Pe^*$. The production rate of filled micelles, i.e., the solubilization rate as well as their diffusivity values are equal in both pure aqueous TTAB solution and in the presence of high M{$_{w}$} PEO. Therefore, the distribution of filled micelles around the moving droplet mainly depends on the droplet propulsive speed, which is faster in the former case. With higher droplet speed the cloud of the filled micelle gets distorted and is more concentrated towards the rear region of the droplet. The distribution of the filled micelles alters the interfacial concentration of the adsorbed surfactants\cite{morozov2020adsorption}. The interfacial surfactant concentration gradient is stronger in the rear hemisphere, where the concentration of filled micelles begins to build-up. This leads to a maximum interfacial flow in the rear region (see FIG. 5A). To satisfy the mass balance, the droplet pulls in fluid from the sides leading to a pusher mode (see FIG. 6A). In the case of high M{$_{w}$} polymer solutions, the droplet speed is significantly lower, resulting in the build-up of the region of filled micelles close to the droplet apex. Such a distribution of filled micelles results in maximum interfacial flow in the frontal hemisphere of the droplet, rendering it as a puller (see FIG. 6B). This is also evident from the PIV measurements of the flow field (FIG. 5B).
\begin{figure}
\centering
\includegraphics[width=1\linewidth]{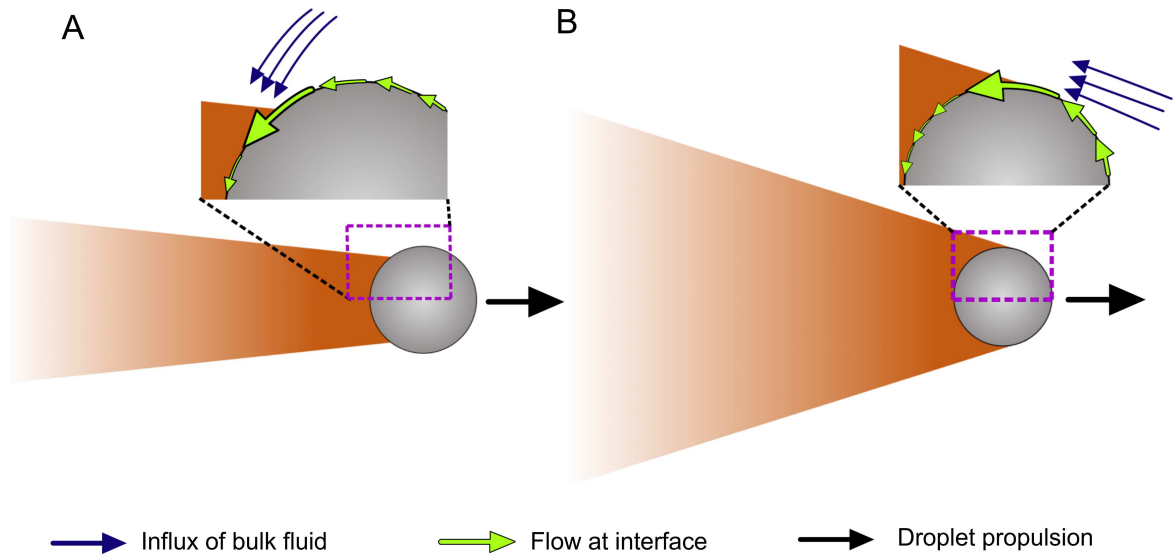}
\caption{Schematic representation of the distribution of filled micelles around the moving droplet. A. Higher concentration gradient of filled micelles with strong interfacial flows at the rear side resulting in an influx of ambient fluid from the sides of the droplet renders it a pusher. B. Higher concentration gradient of filled micelles with strong interfacial flows at the front side resulting in an influx of ambient fluid from the droplet front renders it a puller.}
\label{fig:schematic}
\end{figure} 

Carato \emph{et al.} used a regular perturbation analysis (expansion in $De$) to assess the role of weak viscoelasticity on different squirming modes in a weakly second-order fluid\cite{de2015locomotion}. It showed that in Newtonian fluids, neutral squirmers are the most efficient. However, in weak viscoelastic fluids, pullers dissipate lesser energy and hence are more efficient. In our experiments with no added solutes, the droplet acts as a weak pusher ($\beta \sim$-0.1), which is close to being a neutral swimmer. Further, by adding slight viscoelasticity to the ambient medium (increasing $De$) it transitions to a puller, which is consistent with the predictions of Carato \emph{et al.} 

Finally, exploring the possibility of higher-order transitions (puller to pusher to quadrupole) at higher $Pe^*$ for viscoelastic ambient media, hitherto known only for Newtonian fluids, is also extremely pertinent. So, we investigated the motion of active 5CB droplets in different viscoelastic solutions at higher $c_{TTAB} \sim$15 and 30wt$\%$. The droplets performed smooth motion with pusher mode at 15wt$\%$ and jittery motion with quadrupole mode at 30wt $\%$ (see FIG. 7A). This confirms that even at $De>0$ , increasing $Pe^*$ results in puller to pusher to quadrupole transition. We summarize the mode-switching (puller-to-pusher-to-jittery) in droplet propulsion for the various ambient media considered in this study using a phase diagram plotted between $Pe^{*}$ and $De$ (see FIG. 7B).
\begin{figure}
\centering
\includegraphics[width=1\linewidth]{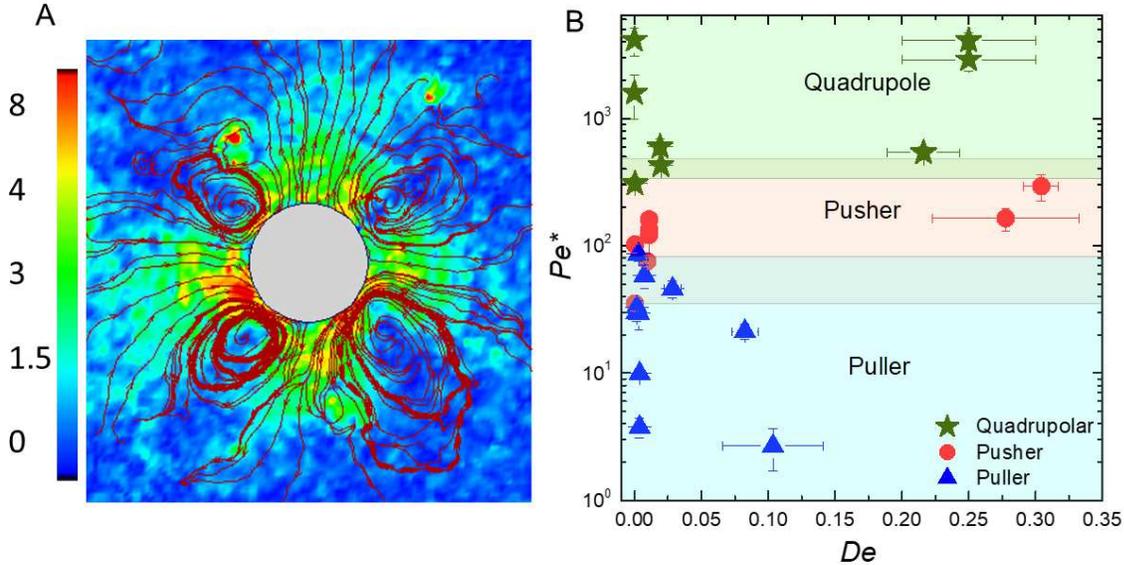}
\caption{A. Fluorescent micrograph obtained from PIV measurements, depicting quadrupolar flow-field around the droplet of size 80$\mu$m B. Phase diagram between $Pe^*$ and $De$, mapping the different modes of droplet swimming under various surroundings.}
\label{phased}
\end{figure} 
Although constructing a detailed phase diagram needs more careful experiments and analytical predictions, our experimental findings strongly suggest that the known based transition of swimming for Newtonian fluids is also observed for viscoelastic solutions, and the addition of viscoelasticity does not alter this transition.

\section*{Conclusion and summary}

We have studied self-propelled motion of droplets in polymer-doped aqueous media. We have demonstrated that by using polymer as macromolecular solutes, the chemical (micelle diffusivity) and hydrodynamic ($\eta_{b}$) field around an active droplet can be tuned independently, by adjusting polymer molecular weight and concentration respectively. By probing the 2D motion of these droplets, we have shown that with increasing molecular weight of the polymers, the droplet trajectories become more persistent. Depending on polymer molecular weight  M{$_{w}$} in polymeric solutions, the micelle diffusivity ($D$) is much faster than the theoretical expectation (SE equation) based on the bulk viscosity, and hence the conventional $Pe$, where $D$ is estimated using $\eta_{b}$ does not accurately predict the mode transitions. A unified framework that utilizes the micelle diffusivity based on the local viscosity of the surrounding environment was demonstrated to capture the transition appropriately. We have also shown that for high molecular weight polymers, upon tuning the polymer and surfactant concentration, the propulsion of active droplets can be selectively manipulated to exhibit different swimming modes i.e. pusher/puller/quadrupolar. We foresee that with synchronous efforts from both experiments and theory/simulations, a detailed phase space can be constructed, which will enhance our understanding of the diverse range of active motions observed in nature.

\section*{Acknowledgements}

This work is supported by the Science and Engineering Research Board (SB/S2/RJN-105/2017), Department of Science and Technology, India.
We also thank Dr. Shashi Thutupalli and Dr. Rajesh Singh for insightlful discussions on our work.

\section*{Author Contributions} PD and RM conceptualized the research, PD, DSP and RM designed the methodology of the experiments, PD performed the experiments, PD, AS and RM analyzed the experimental data, PD, DSP and RM wrote the paper..\\
\section*{Competing Interests} The authors declare that they have no competing financial interests.\\

\bibliography{mybibfile}

\bibliographystyle{BibFileStyle}

\end{document}